\documentclass[12pt,a4paper]{article}
\usepackage{graphicx}
\title{Discrete-time model for a substance motion in a channel of a network. 
Application to  a human migration channel}
\author{Kaloyan N. Vitanov, Nikolay K. Vitanov$^{1,2}$\footnote{corresponding author: vitanov@imbm.bas.bg}}
\date{$^1$Institute of Mechanics, Bulgarian Academy of Sciences, Acad.
G. Bonchev Str., Block 4, 1113 Sofia, Bulgaria \\
$^2$ Max-Planck Institute for the Physics of Complex Systems, Noethnitzerstr.
38, 0187 Dresden, Germany}
\begin{document}
\maketitle
\begin{abstract}
We discuss a discrete-time model for motion of substance in a channel of a network.
For the case of stationary motion of the substance and for
the case of  time-independent values of the parameters of the model we obtain a new class
of statistical distributions  that describe the distribution of the substance
along the nodes of the channel.  The case of interaction between a kind of substance
specific for a node of the network and another kind of substance that is leaked from the channel
is studied in presence of possibility for conversion between the two substances. 
Several scenarios connected to the dynamics of the
two kinds of substances are described. The studied models: (i) model of  motion of substance through a  
channel of a network, and (ii)  model of interaction between two kinds of substances in a network node 
connected to the channel, are discussed from the point of view of human migration dynamics and interaction 
between the population of migrants and the native population of a country.
\end{abstract}
\section{Introduction}
Nonlinear dynamics of complex systems is studied much in the last decades 
\cite{a1} - \cite{vx6} and special attention was set on the areas of social dynamics and population dynamics \cite{albert} - \cite{va15}. 
\par
In the last decades models of flows in networks are much used in the study of different kinds of 
problems, e.g, transportation problems \cite{ff}-\cite{ch1}. In the course of the  years the research 
interest (that initially was  focused on problems such possible maximal flows in a network, minimal cost 
flow problems, or  meeting fixed schedule with minimum number of individuals) expanded to the research 
areas of: just in time scheduling,  shortest path finding, self-organizing network flows,   facility 
layout and location,  modeling and optimization of scalar flows in networks \cite{ambro},  optimal 
electronic route guidance in urban traffic networks \cite{hani}, isoform identification of RNA 
\cite{bernard},  memory effects \cite{rosvall}, etc. (see, e.g., \cite{gomori} - \cite{boz}).
We shall discuss in this article a discrete - time model for the motion of a substance
through a network channel in presence of possibility for "leakage" of substance.  
One possible  application of the discussed model 
is for the flow of  a substance through a channel with use of part 
of the substance in some industrial process in the nodes of the channel. However the model
 has more possible applications and we shall show this for the case of  
human  migration flow. Human migration  is an actual research topic that is  very important for
taking decisions about economic development of regions of a country \cite{everet} -
\cite{borj}.  Human migration is closely connected, e.g., to: (i) 
migration networks \cite{fawcet}, \cite{gurak}; (ii) ideological struggles 
\cite{vit1}, \cite{vit2} ; (iii)  waves and statistical distributions in 
population systems \cite{vit3} - \cite{vit6}. We note that the probability and deterministic
models of human migration are interesting also from the point of view of applied mathematics
\cite{will99} - \cite{grd05}. 
\par 
The text below is organized as follows. In Sect.2 we discuss a discrete - time model for motion of
substance in a channel containing  finite number of nodes. A class of statistical distributions
is obtained in Sect. 3. These distributions describe the distribution of the substance in the nodes of the
channel for the case of stationary motion of substance through the channel.  Particular cases
of the distributions obtained in Sect.3 and in Appendix A are the distributions of  Waring, Yule-Simon, and Zipf . In  Sect. 4 we study the interaction between two kinds of substances in a node of
the network. The substances are: (i) substance that is "native" for the node of the network, and (ii) substance that "leaks" from the corresponding node of the channel to to studied node of the 
network. In Sect. 5 we apply the the results from Sect. 4 to the case of  interaction between population 
of  migrants (the  number of migrants may increase by inflow of migrants from the migration channel) and native population of a country. Several concluding remarks are summarized in Sect. 6. Appendix A contains 
results for the class of statistical distributions that describe the distribution of substance 
along the nodes of the channel for the case of infinite length of the studied channel.
\section{Mathematical formulation of the model}
Let us consider a network consisting of nodes connected by edges.
We assume  existence of a channel in this network - Fig. 1. 
\begin{figure}[!htb]
\centering
\includegraphics[scale=.5]{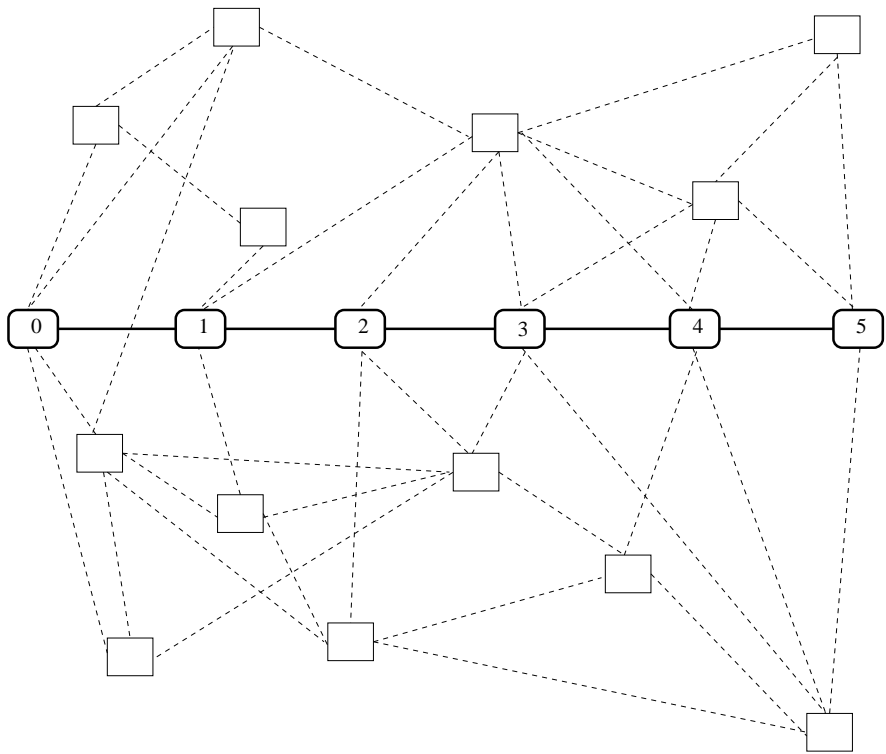}
\caption{A network and a channel. The channel consists of 6 nodes labeled 
from $0$ to $5$. The node $0$ is the entry node of the channel (the substance 
enters the channel through this node). The nodes and the edges  of the channel 
are marked by bold lines. The other nodes and edges of the network are
represented by rectangles and dashed lines.}
\end{figure}
The structure of the channel  and it relation to the network are as follows. 
Several of the nodes of the network together with the corresponding edges 
belong also to the channel.  In Fig. 1 these nodes and edges are marked by bold 
lines. An exchange of substance between the channel 
and the network may happen in the nodes of the channel (denoted as ''leakage" below).
We shall assume that the processes in the network don't influence the flow of the
substance in the channel. The ''leakage'' of the substance from a node of the channel 
however may influence the processes in the corresponding node of the network. 
We shall discuss  such an influence in Sect. 4.
\par
We assume further  that  the channel consists  of a chain 
of $N+1$ nodes (labeled  from $0$ to $N$) connected by corresponding
edges. Each edge connects two nodes and each node is connected to two 
edges except for the $0$-th node and $N$-th node that are connected by 
one edge. We assume that a substance can move through the channel. The substance
enters the channel through the $0$-th node and moves through the channel.
The time is discrete and consists of equal time
intervals. At each time interval the substance in a node of the channel 
(we shall call these nodes also cells below in the text) can participate in
one of the following three processes: (a) the substance remains in the same cell
and stays in the channel; (b) the substance moves to the next node 
(i.e., the substance moves from  the node $m$ to the node $m+1$); 
(c) the substance''leaks'' from the 
channel: this means that the ''leaked'' substance doesn't belong anymore to 
the channel. Such substance may spread through the network. 
In order to obtain intuition about the process of leaking let us
consider a migration channel (this example will be discussed in more detail 
in Sect. 4 and Sect. 5). The network 
in this case is a network of countries connected by roads (e.g., the network
of European countries). The channel consists of several countries connected
by corresponding roads. There is an entry country of the channel and 
there is a last country (sometimes called the final destination country) 
of the channel. What moves in this channel are migrants. They move in the
direction from the entry country of the channel to the final destination
country. In a time interval the migrants: (a) may stay in some of the countries
with an intention to move to the next country of the channel.;
 (b) may move from one country of the
channel to the next country of the channel; or (c) may "leak" from the channel
for some reason (e.g., they may have obtained permission to stay in the
corresponding country of the channel).
\par
Let us formalize mathematically the above considerations. The
 following processes can be observed in a node of the studied channel:
\begin{itemize}
\item
exchange (inflow and outflow) of substance with the previous node of the 
channel (for the nodes $1,\dots,N$-th  of the channel); 
\item
exchange (outflow and inflow) of substance with the next node of the channel 
(for the nodes $0,\dots,N-1$ of the channel;
\item
exchange (inflow and outflow) of substance with the environment of the network;
\item
"leakages":  exchange (outflow and inflow) of substance between the node of the
channel and the the correspondent  node of the network .
\end{itemize}
\par
We consider discrete time $t_k$, $k=0,1,2,\dots$. Let us denote the amount of
the substance in the $i$-th node of the channel at the beginning of the
time interval $[t_k, t_k + \Delta t]$ as $x_i(t_k)$. For the processes happening in this
time interval in the $n$-th node of the channel we shall use the following 
notations 
\begin{itemize}
\item
$i^e_n(t_k)$ and $o^e_n(t_k)$ are the amounts of inflow and outflow  of 
substance from the environment to the $n$-th node  of the channel (the 
upper index $e$ denotes that the quantities are for the environment);  
\item 
$o^c_n(t_k)$ is the amount of outflow of substance from the $n$-th node of the
channel to the $(n+1)$-th node of the channel (the upper index $c$ 
denotes that the quantities are for the channel); 
\item
$i^c_n(t_k)$ is the amount of the inflow of substance from the $(n+1)$ node of 
the channel to the $n$-th node of the channel; 
\item
$o^n_n(t_k)$ and $i^n_n(t_k)$ are the amounts of outflow  
and inflow of substance between the $n$-th node of the channel and  
the corresponding node of the network (the upper index $n$ 
denotes that the quantities are for the network).
\end{itemize}
For the entry node of the channel (the $0$-th node) we have exchange of 
substance with the environment (inflow and outflow); exchange of substance 
with the next node of the channel (inflow and outflow) and "leakage" of 
substance from the channel.
Thus the change  of the amount of substance in the $0$-th node of the channel 
is described
by the relationship
\begin{equation}\label{eq1}
x_0(t_{k+1}) = x_0(t_k)+ i^e_0(t_k) - o^e_0(t_k) - o^c_0(t_k) + i^c_0(t_k) 
-o^n_0(t_k) + i^n_0(t_k)
\end{equation}
For the nodes of the channel  numbered by $i=1,\dots,N-1$ there is exchange
with the environment, "leakage" to the network  and exchange
with $(i-1)$-st and $(i+1)$-st node of the channel. Thus the change  of the amount 
of substance in the $i$-th node of the channel  is described by the relationship
\begin{eqnarray}\label{eq2}
x_i(t_{k+1}) = x_i(t_k)+ i^e_i(t_k) - o^e_i(t_k)  +o^c_{i-1}(t_k)
-i^c_{i-1}(t_k) - o^c_i(t_k) + i^c_i(t_k) - \nonumber \\
o^n_i(t_k) + i^n_i(t_k) , \ \ \ 
i=1,\dots,N-1
\end{eqnarray}
For the last node (the $N$-th node of the channel) there is exchange with the
environment, "leakage" to the network and exchange  
with $(N-1)$-st node of the channel. Thus the change  of the amount of 
substance in the $i$-th node  of the channel  is described by the relationship
\begin{eqnarray}\label{eq3}
x_N(t_{k+1}) = x_N(t_k)+ i^e_N(t_k) - o^e_N(t_k)  + o^c_{N-1}(t_k)
-i^c_{N-1}(t_k) - \nonumber \\
o^n_N(t_k) + i^n_N(t_k) 
\end{eqnarray}
\par
Eqs.(\ref{eq1}) - (\ref{eq3}) describe the general case of motion of substance
along the channel of the studied network. Below we shall discuss a particular
case where no exchange with the environment is present except for the entry
node of the channel.  In addition we shall assume that:
\begin{itemize}
\item
 there is no inflow of 
substance from the nodes of the network to the channel. 
\item
there is no outflow of substance from the $0$-th node of the channel to the
environment
\item
there is no inflow of substance from the $i$-th node of the channel to the
$i-1$-th node of the channel, $i=1,\dots,N$.
\end{itemize}
For the particular case described above the system of model equations 
(\ref{eq1}) - (\ref{eq3}) becomes
\begin{equation}\label{e1}
x_0(t_{k+1}) = x_0(t_k)+ i^e_0(t_k)  - o^c_0(t_k)  -o^n_0(t_k)
\end{equation}
\begin{eqnarray}\label{e2}
x_i(t_{k+1}) = x_i(t_k)+o^c_{i-1}(t_k) - o^c_i(t_k)  - o^n_i(t_k), \ \   
i=1,\dots,N-1
\end{eqnarray}
\begin{eqnarray}\label{e3}
x_N(t_{k+1}) = x_N(t_k) + o^c_{N-1}(t_k) - o^n_N(t_k) 
\end{eqnarray}
\par
Below we shall study the following particular cases of the quantities from
the system of equations (\ref{e1}) - (\ref{e3})
\begin{eqnarray}\label{pc}
i^e_0(t_k) &=& \sigma(t_k) x_0(t_k); \ \ o^c_0(t_k) = f_0(t_k) x_0(t_k);
\nonumber \\
o^n_0(t_k) &=& \gamma_0(t_k) x_0(t_k); \ \ o^c_{i-1}(t_k) = f_{i-1}(t_k)
x_{i-1}(t_k);
\nonumber \\
o^c_{i}(t_k) &=& f_{i}(t_k) x_i(t_k); \ \ o^n_{i}(t_k) = \gamma_{i}(t_k)
x_i(t_k);
\nonumber \\
o^c_{N-1}(t_k) &=& f_{N-1}(t_k) x_{N-1}(t_k); \ \ o^n_{N}(t_k) = \gamma_{N}(t_k)
x_N(t_k);
\end{eqnarray}
For this particular case the system of equation (\ref{e1}) - (\ref{e3}) becomes
\begin{equation}\label{ex1}
x_0(t_{k+1}) = x_0(t_k) + \sigma(t_k) x_0(t_k) - f_0(t_k) x_0(t_k) - 
\gamma_0(t_k) x_0(t_k)
\end{equation}
\begin{equation}\label{ex2}
x_i(t_{k+1}) = x_i(t_k) + f_{i-1}(t_k)x_{i-1}(t_k) - f_{i}(t_k) x_i(t_k) - \gamma_{i}(t_k) x_i(t_k) \ \   
i=1,\dots,N-1
\end{equation}
\begin{equation}\label{ex3}
x_N(t_{k+1}) = x_N(t_k) + f_{N-1}(t_k) x_{N-1}(t_k) - \gamma_{N}(t_k)
x_N(t_k)
\end{equation}
We shall study the model equations (\ref{ex1}) - (\ref{ex3}) in more detail
below.
\section{Distributions of substance corresponding to stationary regime 
of functioning of the channel }
Below we discuss the model described by Eqs.(\ref{ex1}) -
(\ref{ex3}) for the case when the parameters of the model are time independent
(i.e., when $\sigma(t_k) = \sigma$; $\alpha_i(t_k) = \alpha_i$, $i=0,\dots,N$;
$\gamma_i(t_k) = \gamma_i$,$i=0,\dots,N$, $f_i(t_k) = f_i$,$i=0,\dots,N$ ).
In this case the system of model equations becomes
\begin{equation}\label{ey1}
x_0(t_{k+1}) = x_0(t_k) + \sigma x_0(t_k) - f_0 x_0(t_k) - 
\gamma_0 x_0(t_k)
\end{equation}
\begin{equation}\label{ey2}
x_i(t_{k+1}) = x_i(t_k) + f_{i-1}x_{i-1}(t_k) - f_{i} x_i(t_k) - \gamma_{i}x_i(t_k) \ \   
i=1,\dots,N-1
\end{equation}
\begin{equation}\label{ey3}
x_N(t_{k+1}) = x_N(t_k) + f_{N-1} x_{N-1}(t_k) - \gamma_{N}
x_N(t_k)
\end{equation}
\par
In addition we shall consider the stationary state: $x_i(t_k) = x_i^*$. This
stationary state occurs when $x_i(t_{k+1})=x_i(t_k)$ (i.e., there
is a motion of substance through the cells of the channel but the motion
happens in such a way that the amount of the substance in a given cell
remains the same in the course of the time). From the system of
equations (\ref{ey1}) - (\ref{ey3}) we obtain ($i=1,\dots,N-1$, $x_0^*$ is a
free parameter)
\begin{eqnarray}\label{dstr}
x_i^* = x_0^* \prod \limits_{j=1}^i \frac{f_{j-1}}{f_j+\gamma_j}; \ \
x_N^* = x_0^* \frac{f_{N-1}}{\gamma_N}\prod \limits_{j=1}^{N-1} \frac{f_{j-1}}{f_j+\gamma_j}
\end{eqnarray}
The total amount of the substance in the channel is
\begin{equation}\label{tas}
x^* = x_0^* \Bigg[ 1 + \sum \limits_{k=1}^{N-1} \prod \limits_{j=1}^k \frac{f_{j-1}}{f_j+\gamma_j}
+ \frac{f_{N-1}}{\gamma_N}\prod \limits_{j=1}^{N-1} \frac{f_{j-1}}{f_j+\gamma_j}
\Bigg]
\end{equation}
We can consider the statistical distribution $y^*_i=x_i^*/x^*$ of the amount of substance along
the nodes of the channel. $y_i^*$ can be considered as probability values
of distribution of a discrete random variable $\zeta$: $y_i^* = p(\zeta =i)$, $i=1,\dots, N$.
For this distribution we obtain
\begin{eqnarray}\label{dstr_main1}
y^*_0 &=& \frac{1}{\Bigg[ 1 + \sum \limits_{k=1}^{N-1} \prod \limits_{j=1}^k \frac{f_{j-1}}{f_j+\gamma_j}
+ \frac{f_{N-1}}{\gamma_N}\prod \limits_{j=1}^{N-1} \frac{f_{j-1}}{f_j+\gamma_j}
\Bigg]} \nonumber \\
y_i^* &=& \frac{\prod \limits_{j=1}^i \frac{f_{j-1}}{f_j+\gamma_j}}{\Bigg[ 1 + \sum \limits_{k=1}^{N-1} \prod \limits_{j=1}^k \frac{f_{j-1}}{f_j+\gamma_j}
+ \frac{f_{N-1}}{\gamma_N}\prod \limits_{j=1}^{N-1} \frac{f_{j-1}}{f_j+\gamma_j}
\Bigg]}; i=1,\dots,N-1 \nonumber \\
y_N^* &=& \frac{\frac{f_{N-1}}{\gamma_N}\prod \limits_{j=1}^{N-1}
\frac{f_{j-1}}{f_j+\gamma_j}}{\Bigg[ 1 + \sum \limits_{k=1}^{N-1} \prod \limits_{j=1}^k \frac{f_{j-1}}{f_j+\gamma_j}
+ \frac{f_{N-1}}{\gamma_N}\prod \limits_{j=1}^{N-1} \frac{f_{j-1}}{f_j+\gamma_j}
\Bigg]}
\end{eqnarray}
Eq.(\ref{dstr_main1}) describes a class of statistical distributions ($f_i$ and
$\gamma_i$ are still not specified). To the best of our knowledge the general form
(\ref{dstr_main1}) of this class  of 
distributions was not discussed by other authors.
\begin{figure}[!htb]
	\centering
	\includegraphics[scale=.7]{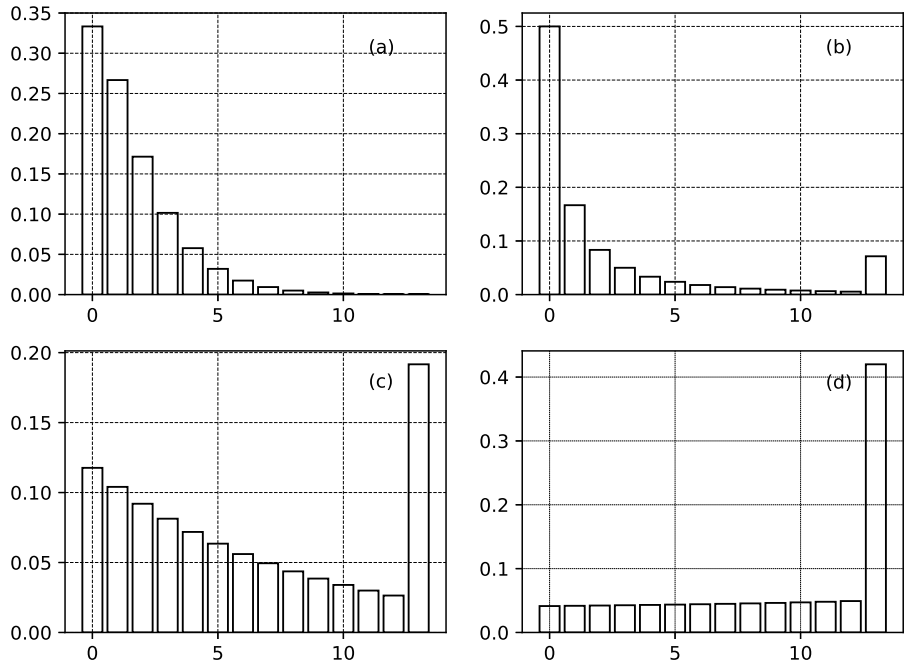}
	\caption{ Several examples for  distributions from the class of distributions  (\ref{dstr_main1}) for a 
	channel consisting of 14 nodes. Figure (a): 	$\gamma_i=0.001$,  $f_i=0.001/(i+1)+0.001$. Figure (b): $\gamma_i=0.001$,
	$f_i=0.001(i+1)$. Figure (c) $\gamma_i=0.001$,
	$f_i=0.0075 - 0.00002i$. Figure (d): $\gamma_i=0.00041$,  $f_i=0.01 - 0.0005(i+1)$.
}
\end{figure}
Fig. 2 shows several examples for distributions of the class (\ref{dstr_main1}).  
The form of the distribution can be standard as in Fig. 2a but there exist another possible form connected to
concentration of substance in the last node of the channel - Figs. 2b,c,d. Fig.2d shows a an interesting form
of the distribution that can arise only in channel having finite number of nodes: the probability increases with increasing number of the node. We shall discuss again the distributions from Fig. 2 in Sect. 5 where we shall 
consider the application of the model to channels of human migration.
\par
 We 
note that the class of distributions (\ref{dstr_main1}) has interesting
particular cases that have been discussed in connection with channels of
migration of substance or migration channels of human migration. For and
example let $f_i=\alpha_i+\beta_i i$,
$i=1,\dots,N$, $\alpha_i>0$, $\beta_i \ge 0$, $\sigma_0 >0$, $\gamma_i \ge 0$. 
Then the stationary amount $x_i^*$ of the substance along the modes of the
channel is given by the relationship
\begin{eqnarray}\label{distr1x}
x_i^* &=& \frac{\prod \limits_{j=1}^i [\alpha_{i-j} + (i-j) \beta_{i-j}] }{
\prod \limits_{j=1}^i (\alpha_j + j \beta_j + \gamma_j)} x_0^*, \
i=1,\dots, N-1 \nonumber\\
x_N^* &=& \frac{\prod \limits_{j=1}^N [\alpha_{N-j} + (N-j) \beta_{N-j}]}{ \gamma_N \prod \limits_{j=1}^{N-1} (\alpha_j + j \beta_j + \gamma_j)} x_0^*
\end{eqnarray}
and the statistical distribution connected to this stationary state of
functioning of the channel is
 \begin{eqnarray}\label{distr2x}
y_0^* &=& \frac{1}{1+  \sum\limits_{i=1}^{N-1}\frac{\prod \limits_{j=1}^i [\alpha_{i-j} + (i-j) \beta_{i-j}] }{
\prod \limits_{j=1}^i (\alpha_j + j \beta_j + \gamma_j)} + \frac{\prod \limits_{j=1}^N [\alpha_{N-j} + 
(N-j) \beta_{N-j}]}{ \gamma_N \prod \limits_{j=1}^{N-1} (\alpha_j + j \beta_j + 
\gamma_j)}} \nonumber \\  
y_i^* &=& \frac{\frac{\prod \limits_{j=1}^i [\alpha_{i-j} + (i-j) \beta_{i-j}] }{
\prod \limits_{j=1}^i (\alpha_j + j \beta_j + \gamma_j)}}{1+  \sum\limits_{i=1}^{N-1}    \frac{\prod \limits_{j=1}^i [\alpha_{i-j} + (i-j) \beta_{i-j}] }{
\prod \limits_{j=1}^i (\alpha_j + j \beta_j + \gamma_j)} + \frac{\prod \limits_{j=1}^N [\alpha_{N-j} + 
(N-j) \beta_{N-j}]}{ \gamma_N \prod \limits_{j=1}^{N-1} (\alpha_j + j \beta_j + 
\gamma_j)}}, \
i=1,\dots, N-1 \nonumber\\
y_N^* &=& \frac{\frac{\prod \limits_{j=1}^N [\alpha_{N-j} + 
(N-j) \beta_{N-j}]}{ \gamma_N \prod \limits_{j=1}^{N-1} (\alpha_j + j \beta_j 
+ \gamma_j)}}{1+\sum\limits_{i=1}^{N-1} \frac{\prod \limits_{j=1}^i [\alpha_{i-j} + (i-j) \beta_{i-j}] }{
\prod \limits_{j=1}^i (\alpha_j + j \beta_j + \gamma_j)} + \frac{\prod \limits_{j=1}^N [\alpha_{N-j} + 
(N-j) \beta_{N-j}]}{ \gamma_N \prod \limits_{j=1}^{N-1} (\alpha_j + j \beta_j + 
\gamma_j)}}
\end{eqnarray}
The distribution (\ref{distr2x}) is a generalization, e.g., of the truncated Waring distribution \cite{vk} as well
 as a generalization of one of distributions discussed in \cite{vk1}.  
We note that the corresponding distribution for the case of channel of infinite length is a
generalization of the Waring distribution and because of this it contains as
particular cases several famous distributions such as Zipf distribution or
Simon-Yule distribution (see Appendix A). 
\begin{figure}[!htb]
	\centering
	\includegraphics[scale=.7]{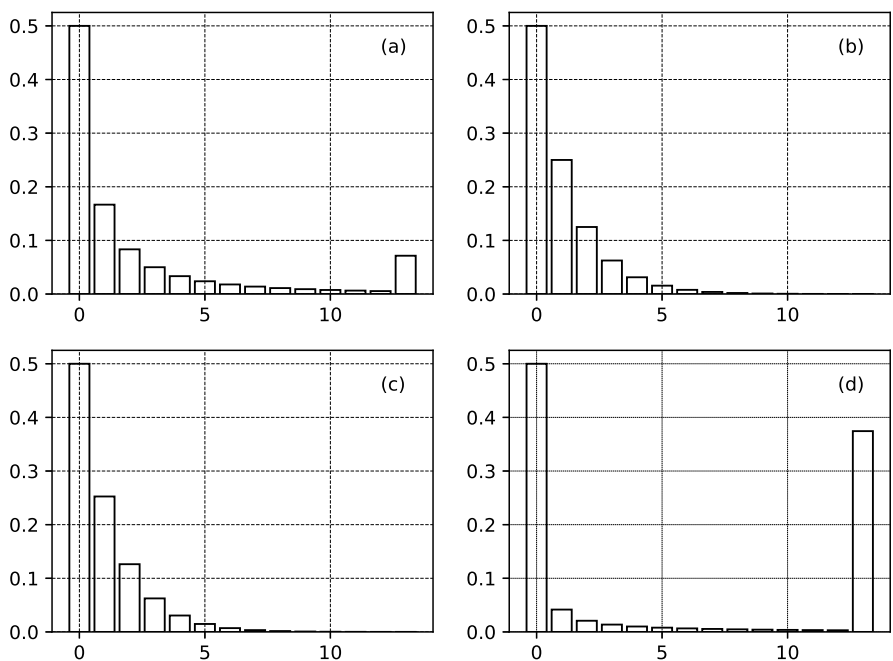}
	\caption{ The distribution (\ref{distr2x}) for a channel consisting of 14 nodes.
All parameters $\gamma_i$ have the same value $\gamma_i=0.001$ ($i=0,\dots,13$) in all
figures - 3a, 3b, 3c, 3d. All parameters $\alpha_i$ have the same value $\alpha_i=0.001$ ($i=0,\dots,13$)  in all
figures - 3a, 3b, 3c, 3d. Figures show the changes in the form of the distribution when the parameter $\beta_i$
is changed.  Figure (a): $\beta_i = 0.001i$. There is a concentration of the substance in the last node of the channel.
Figure (b): $\beta_i=0$. The amount of substance in the second half of the channel decreases in comparison to
Fig. (a). There is no concentration of substance in the last node of the channel. Figure (c): $\beta_i=-0.00002i$.
Amount of substance in the second half of the channel decreases in comparison to the case from Fig. (b).
Figure (d): $\beta_i = 0.01i$.  There is a large concentration of substance in the last node of the channel.}
\end{figure}
The distribution (\ref{distr2x}) is visualized in Fig. 3 for fixed values of the parameters $\alpha_i$ and $\gamma_i$ 
and for different values of the parameters $\beta_i$. As we can see the values of the parameters $\beta_i$ 
influence the situation about the amount of the substance in the second half of the channel.  Specific feature of the discussed distributions is the relatively large value of the probability  (large value of the amount of substance) 
in the last node of the channel - Fig. 3a.
This probability can increase if the values of $\beta_i$ are increased - Fig. 3d. The probability can decrease when the values of $\beta_i$ are set to $0$ or become negative - Figs. 3b, 3c. Then the tendency is for concentration of substance in
the first half of the channel. We shall discuss the distribution (\ref{distr2x}) below in the text as it describes an interesting
situation for the case of channel of human migration, namely the situation where the attractiveness of the countries from the
second half of the channel is larger (positive values of $\beta_i$) (or smaller - negative values of $\beta_i$) with respect to the
attractiveness of the countries from the first half of the channel.
Let us note that more information about the corresponding distribution for the case of
infinite channel can be obtained from Appendix A.
\section{On interaction between substances in  a node  that belongs to the channel and  to the network}
As we have seen above a part of the substance may leave the channel as an
outflow  (''leakage") from the channel to the corresponding node of the network.
Let us study the following problem. We consider a node  of the network
that contains some amount $A(t_k)$ of substance $A$ and obtains (through its
connection with the channel) amount $c(t_k)$ of the substance $B$ in the time
interval between $t_k$ and $t_{k+1}$. Let us assume presence of  two kinds 
of  processes in the discussed node. These processes can: (i) lead to change of the amount of substances $A$
and $B$ without conversion of $A$ to $B$  and $B$ to $A$, and (ii) lead to
changes in the amount of the substances $A$ to $B$ by means of conversion   
of $A$ to $B$ and $B$ to $A$. The model system for the change of the amount of
substances in the discussed node will be
\begin{eqnarray}\label{dem1}
A(t_{k+1})&=& A(t_k)+ p(t_k)A(t_k)+q(t_k)A(t_k)B(t_k), \nonumber \\
B(t_{k+1})&=& B(t_k)+ c(t_k) + r(t_k)B(t_k) -q(t_k)A(t_k)B(t_k),
\end{eqnarray}
where $p(t_k)$ is the parameter that describes the changes in the amount of the
substance $A$ as a result of processes that do not lead to conversion between
$A$ and $B$; $r(t_k)$ is the parameter that describes the changes in the amount of the
substance $B$ as a result of processes that do not lead to conversion between
$A$ and $B$; $q(t_k)$ is parameter that describes the changes as a
consequence of the conversion between $A$ and $B$. The stationary state
of the system (\ref{dem1}) ($A(t_{k+1})= A(t_k)$; $B(t_{k+1})= B(t_k)$) is
\begin{equation}\label{dem2}
A^*(t_k)= \frac{r(t_k)p(t_k)-c(t_k)q(t_k)}{p(t_k)q(t_k)}; \ \ 
B^*(t_k)= - \frac{p(t_k)}{q(t_k)} 
\end{equation}
\begin{figure}[!htb]
\centering
\includegraphics[scale=.7]{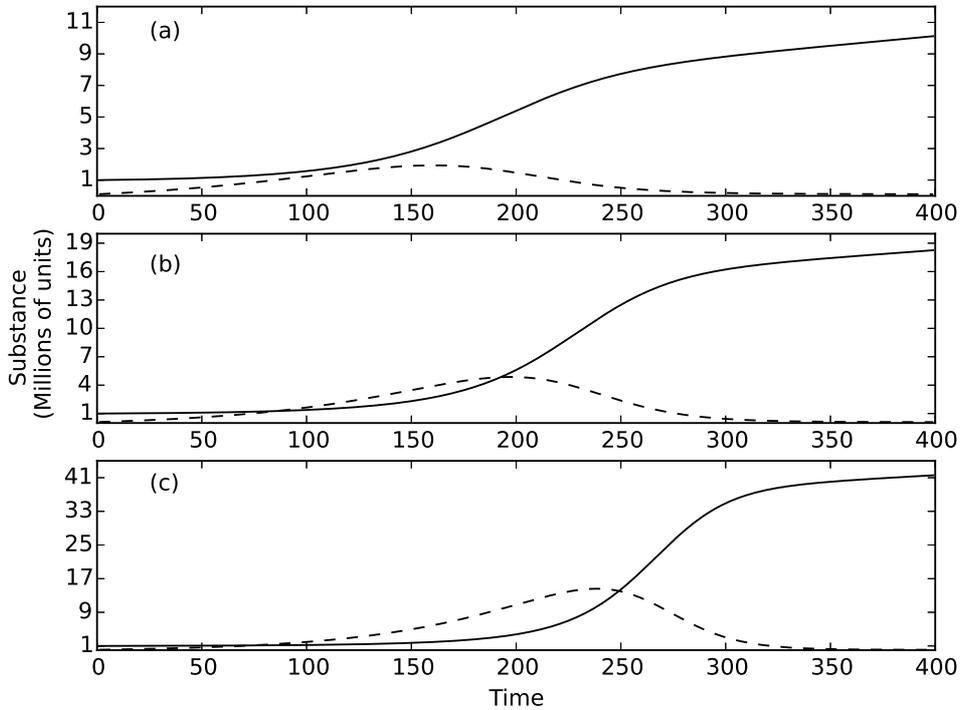}
\caption{Influence of parameter $q$ for the case of constant values of the
parameters in Eqs. (\ref{dem1}). Solid line: amount of substance $A$.
Dashed line: amount of substance $B$. 
Figure (a): $q=7 \cdot 10^{-9}$. Figure (b):
$q=4 \cdot 10^{-9}$. Figure (c): $q=2 \cdot 10^{-9}$. The values of 
the other parameters are: $p=5 \cdot 10^{-4}$, $r=2 \cdot 10^{-2}$, $c=5 \cdot
10^{3}$. The initial conditions  are: $A(0)=10^6$, $B(0)=10^5$.}
\end{figure}
Let us first  assume that the parameters $c$,$p$,$q$,$r$ don't depend on time.
Fig. 4 shows the influence of the parameter $q$ of the amounts of the substances
and describes \emph{Scenario No.1: Limitation of the amount of substance $B$ by conversion}.
Larger values of $q$ mean that larger amount of the substance $B$ is
converted to substance $A$. The other parameters are chosen in such a way that
the rate of increase of substance $B$ is larger than the rate of increase
of the substance $A$ and in addition the substance $B$ increases also by means
of some substance that arrives at the node from the the channel. Despite the favorable
conditions for increase of the amount of substance $B$  the presence of possibility 
for conversion of $B$ to $A$ leads to a result that $B$ decreases to very low values after
some time - Fig. 4a. Even if the conversion rate decreases and even if the
amount of substance $B$ becomes larger than the amount of substance $A$ (Fig. 
4b) the final result may be the same. Further decreasing of the value of $q$
may lead to large time of dominance of the substance $B$ in the node of the
network - Fig. 4c but the final result can be the same as in the other two
figures:  because of the conversion $A$ prevails and $B$ is reduced to
negligible amounts. Additional decreasing of the value of $q$ can however lead to
change of the situation. At some critical low value of $q$ the conversion can't
compensate anymore the rate of increase of $B$ and the amount of $B$ can exceed $A$.
This dominance of the amount of substance $B$ can last as long as the parameters 
of the system in the studied node of the system remain unchanged. Such a situation 
can be easily observed if we just put $q=0$ in Eqs.(\ref{dem1}). There is no more 
coupling of the amounts of the substances $A$ and $B$ and because of its larger rate of
increasing  it is just a matter of time for the amount of substance $B$ to exceed
the amount of substance $A$. All this shows that the mechanism of  limitation by 
conversion has its limits.
\begin{figure}[!htb]
\centering
\includegraphics[scale=.7]{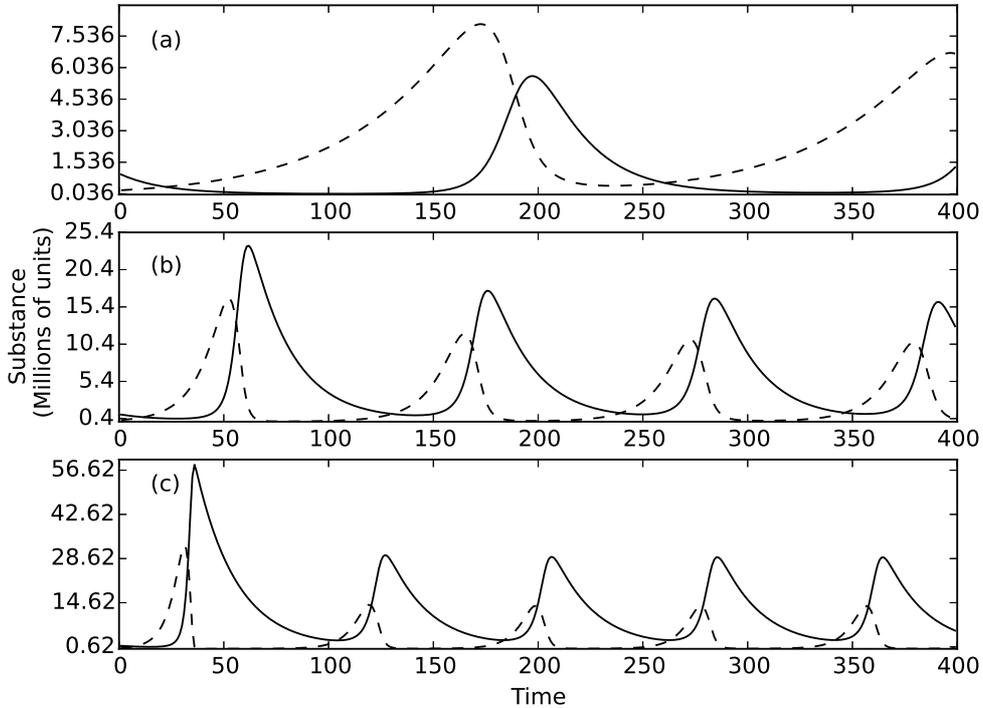}
\caption{ Interplay of conversion and negative rate $p$ of increase of the 
substance $A$ with the influence of increasing positive rate $r$ of increase of
the amount of the substance $B$. Solid line: amount of substance $A$.
Dashed line: amount of substance $B$. Initial conditions are: $A(0)=10^6$,
$B(0)=2\cdot10^5$. The values of the parameters are as
follows. $p=-0.05$, $q=2\cdot10^{-8}$, $c=5000$.
Figure (a): $r=0.02$. Figure (b): $r=0.1$. Figure (c): $r=0.2$. The rate of (non-conversion) 
increase of the amount of substance $B$ influences the cyclic behavior of the
substances $A$ and $B$.}
\end{figure}
\par
The combination of appropriate values of parameters can lead to interesting
evolution of the amounts of the the substances $A$ and $B$ in the studied node
of the network. Fig. 5  shows the influence of negative rate of
increase of the substance $A$ compensated by positive rate of conversion from
substance $B$ to substance $A$. The scenario here is \emph{Scenario No.2: Cyclic behavior of the
amounts of the substance $A$ and $B$}.
The value of the rate  of increase $r$ of substance $B$ increases
from Fig. 5a to Fig. 5c. The result is a cyclic evolution of the amounts
of substances $A$ and $B$ and the period of the cycle is influenced by the
parameter $r$: an increase of $r$ leads to a decrease of the value of the period.
\begin{figure}[!htb]
\centering
\includegraphics[scale=.7]{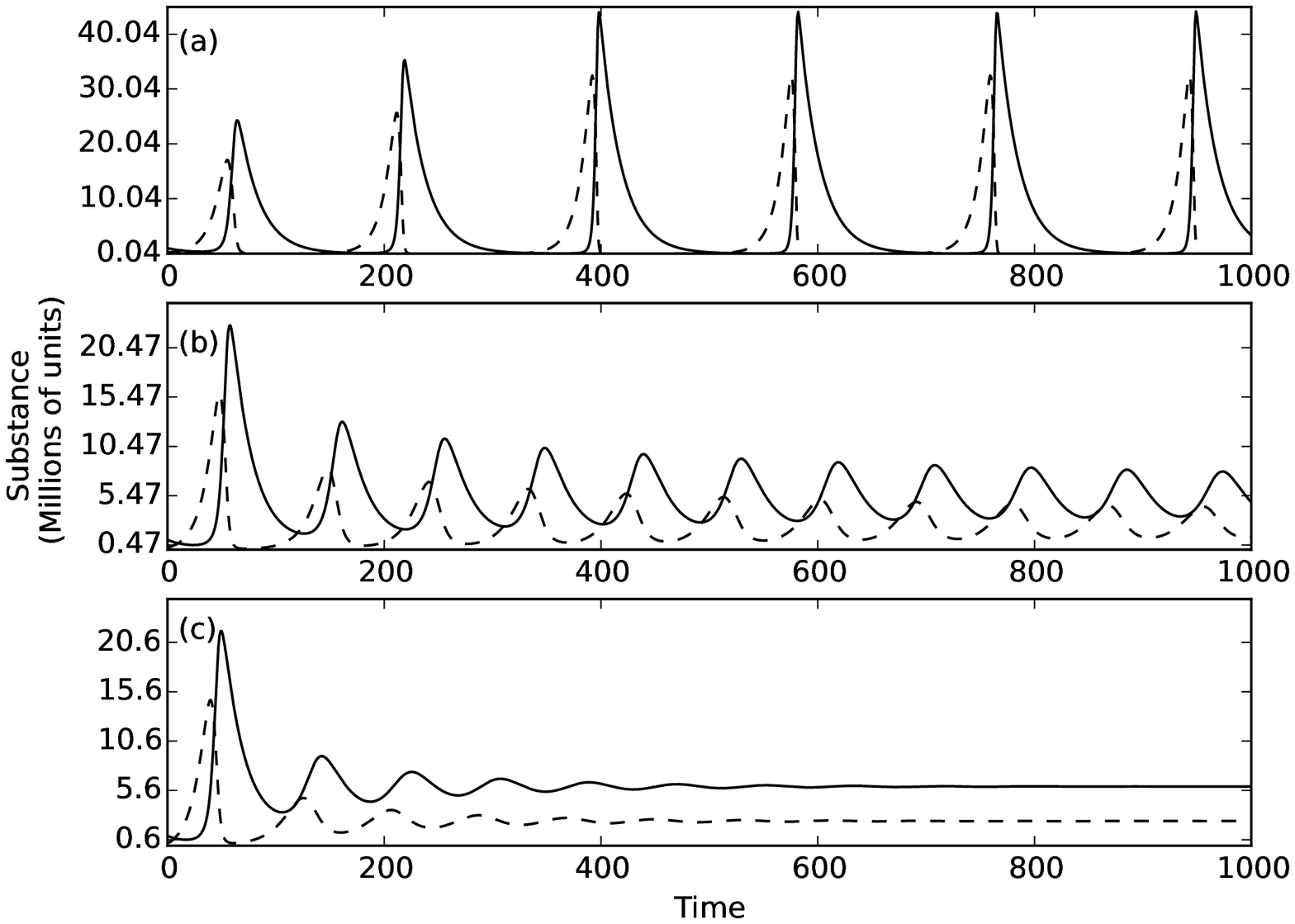}
\caption{Influence of increasing value of parameter $c$ on the dynamics of
the amounts of substances $A$ and $B$ in the studied node of the network .
Interplay of conversion and negative rate of increase of the 
substance $A$ with the influence of increasing positive rate of increase of
the amount of the substance $B$. Solid line: amount of substance $A$.
Dashed line: amount of substance $B$. Initial conditions are: $A(0)=10^6$,
$B(0)=2\cdot10^5$. The values of the parameters are as
follows. $p=-0.05$, $q=2\cdot10^{-8}$, $r=0.1$.
Figure (a): $c=20$. Figure (b): $c=15,000$. Figure (c): $c=50,000$.}
\end{figure}
\par
The parameter $c$ (it regulates the inflow of substance $B$ from the
the channel to the studied node of the network) can have considerable influence
on the dynamics of the amounts of substances $A$ and $B$ in the studied node
of the network. One scenario: \emph{Scenario No.3: Dominance through inflow  
and conversion of substance} connected to such a large influence is shown in
Fig.6.  Fig. 6a shows the situation in presence of a negligible
amount of substance $B$ coming from the channel. The inflow of substance $B$
in not felt and the main processes that determine the dynamics of the substances
in the node are the conversion of $B$ to $A$, the decrease of $A$ (e.g., because
of its use in some process) and the increase of the amount of $B$ due to
processes happening inside the studied node of the network. As we can see a
cyclic behavior of the amounts of substances occurs in the studied node. Let us
now begin to increase the parameter $c$, i.e., the amount of substance $B$
per unit time increases that flows in  the studied node from the channel. The
characteristics of the cyclic behavior of the amounts of the substances in the
studied node change. In Fig 6b the value of $c$ is 750 time larger that the
value of $c$ for Fig 6a. The period of the observed cycle decreases. The
explanation is that the fast increasing of the substance $B$ activates
the processes that convert substance $B$ to substance $A$ and these processes
are much more intensive in comparison to the case shown in Fig. 6a. Further
increasing of the value of $c$ (further increasing of the amount of inflow
of substance $B$ from the channel to the node of the network) leads to vanishing
of the cyclic behavior - Figure 6c. Instead of this a stationary state occurs where
the amount of the substances $A$ and $B$ in the studied node have constant
values. 
\begin{figure}[!htb]
\centering
\includegraphics[scale=.7]{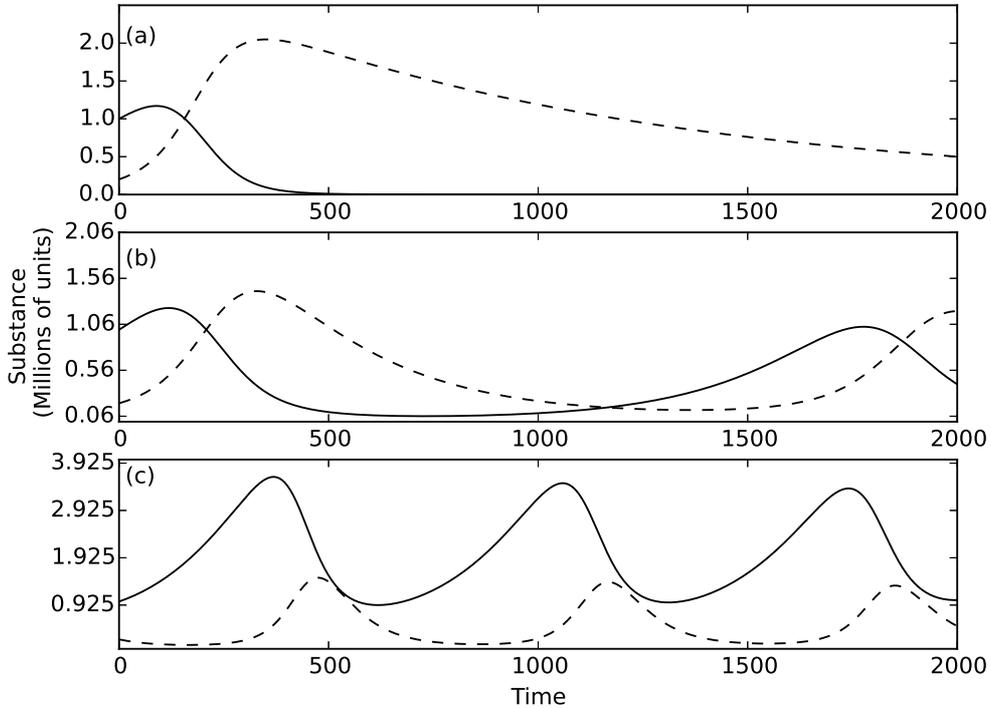}
\caption{Influence of negative value of parameter $r$  and negative value of
parameter $q$ on the dynamics of the amounts of substance in the studied node 
of the network. Solid line: amount of substance $A$.
Dashed line: amount of substance $B$. Initial conditions are: $A(0)=10^6$,
$B(0)=2\cdot10^5$. The values of the parameters are as
follows. $p=0.005$, $q=-10^{-8}$, $c=100$.
Figure (a): $r=-0.001$. Figure (b): $r=-0.004$. Figure (c): $r=-0.02$.
}
\end{figure}
\par
The last situation we shall discuss for the case of constant values of the
parameters in the model equations is the situation of positive $p$ (the processes
in the node of the network lead to increasing amount of the substance
$A$), negative $r$  (the processes in the node of the network lead to 
decreasing  amount of the substance $B$) and negative values of $q$ (processes
happen in the network node that lead to conversion of substance $A$ to
substance $B$)  - \emph{Scenario No.4: Conversion can't compensate for decreasing}. 
Fig.7 shows the influence of decreasing values of the
rate $r$. Large values of $r$ lead to dominance of the substance $B$: despite
the smaller value of this substance at the node of the network in the initial
moment of time the amount of the substance $B$ increases fast  because of the 
conversion and then remains larger that the amount of the substance
$A$ (and this is despite the fact that the rate $p$ is positive). The decreasing
of the value of the rate $r$ leads to appearing of cyclic behavior and there is
an interval of values of the rate $r$ where the dominance is exchanged: for some
time interval the amount of the substance $B$ is larger than the amount of the
substance $A$ and in the next time interval the amount of the substance $A$ is
larger that the amount of the substance $B$. If the value of the ratio $r$
decreases further then the cyclic behavior in the node may persist for the long
time but the substance $A$ remains dominant. At some value of $r$ the cyclic
behavior vanishes. The vanishing may take some time (as in the case of Fig. 6c).
or may be faster if the value of $r$ is small enough.
\begin{figure}[!htb]
\centering
\includegraphics[scale=.7]{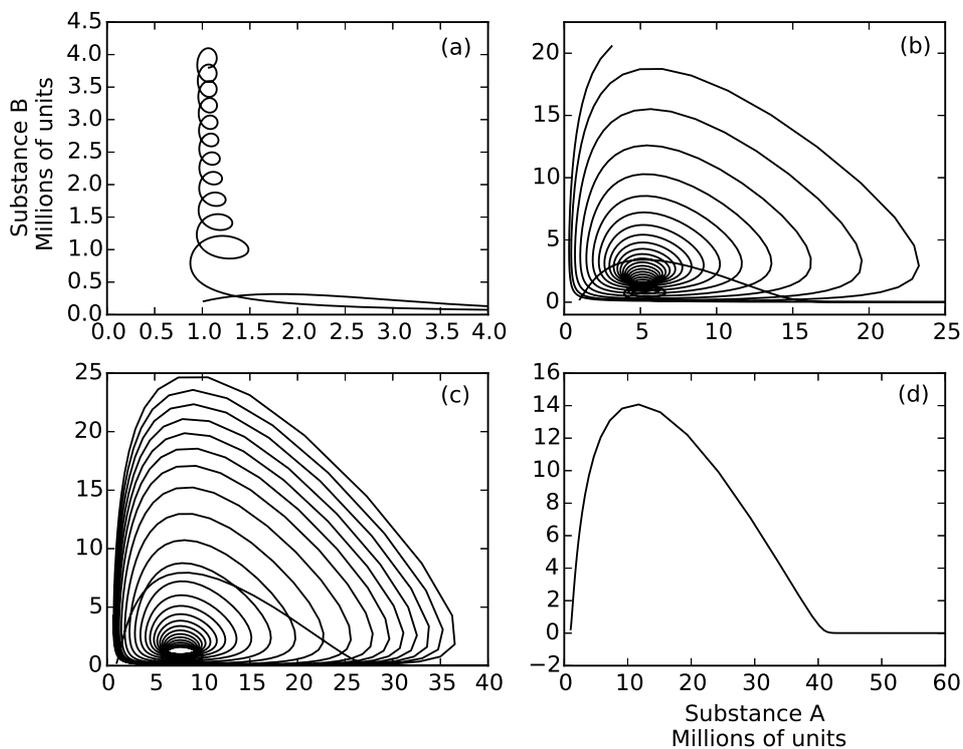}
\caption{ Evolution of the system for the case of coefficient $p(t_k)$ that
decreases with increasing time. On all figures $p(t_k)=0.01-0.00003 t_k$. The
other parameters of the system (\ref{dem1}) are: $q=2 \cdot 10^{-8}$,
$c=5000$. Parameter $r$ has different vales as follows. Figure (a): $r=0.02$.
Figure (b): $r=0.1$. Figure (c): $r=0.15$. Figure (d): $r=0.2$ The initial
values for the amounts of the substance are: $A(0)=10^6$, $B(0)=2\cdot 10^5$.
}
\end{figure}
\par
Let us now discuss Eqs.(\ref{dem1}) for some cases when the participating
parameters change their values in the course of the time. If Figs. 8 and 9 we
present results for such situation in which the parameter $p$ decreases slowly
with the time for some finite interval of time. 
This decrease is the same for Figs. 8a - 8d. In addition we have  different
values of parameter $r$:  the smallest  value of $r$ is for the situation shown in Fig.8a
and the largest value of $r$ is  for the situation shown in Fig. 8d. Two oscillation
regimes are observed. Let us call them regime of smaller amplitude oscillations
and regime of larger amplitude oscillations. The regime of smaller amplitude
oscillations is shown in Fig. 8a. The specific feature of this regime is
that the amount of the substance $A$ oscillates in time around a fixed value
whereas the amount of the substance $B$ oscillates in time around a line
characterizing a trend of increase of the value of substance $B$. The increasing
value of parameter $r$ decreases the interval of time in which the substances 
from the network node change their amount in the regime of smaller amplitude
oscillations. This regime is followed by a regime of larger amplitude
oscillations - Fig. 8b.  Further increase of the value of $r$ leads to an 
increase of the time in which the system of two substances is in the regime of 
larger amplitude oscillations. Finally the increase of the value of $r$
above some threshold value leads to vanishing of the oscillation regime for
the values of the substances - Fig. 8d.
\begin{figure}[!htb]
\centering
\includegraphics[scale=.7]{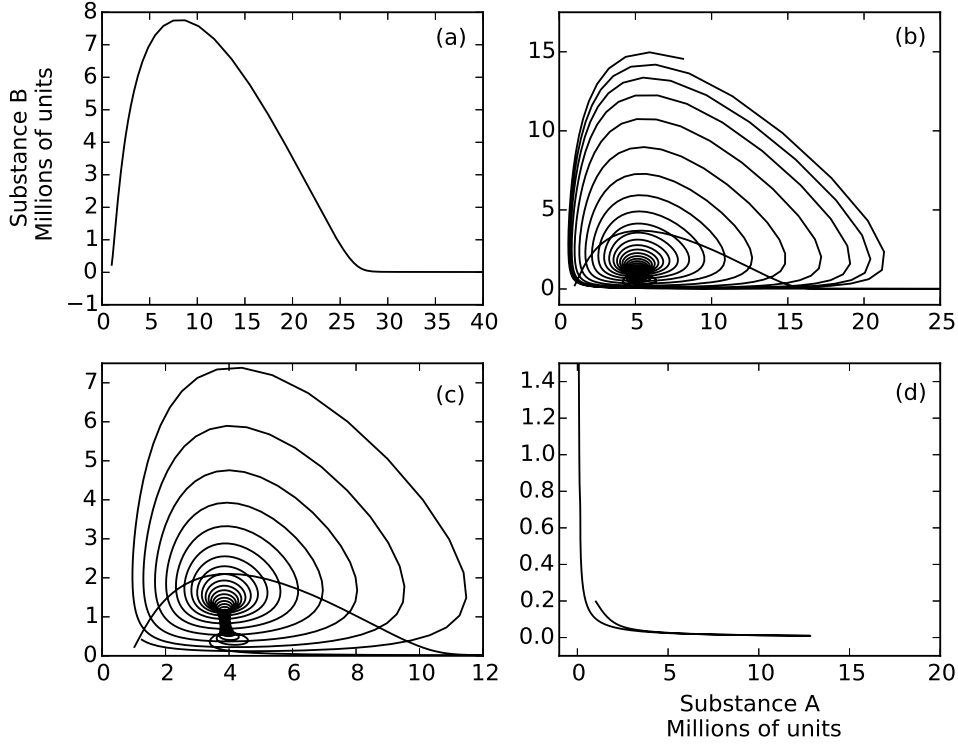}
\caption{
Evolution of the system for the case of coefficient $p(t_k)$ that
decreases with increasing time. On all figures $p(t_k)=0.011-0.00003 t_k$. The
other parameters of the system (\ref{dem1}) are: $r=0.15$,
$c=5000$. Parameter $q$ has different vales as follows. Figure (a):
$q=2 \cdot 10^{-8}$ .
Figure (b): $q=3 \cdot 10^{-8}$. Figure (c): $q=4 \cdot 10^{-8}$. 
Figure (d): $q=5 \cdot 10^{-8}$. The initial
values for the amounts of the substance are: $A(0)=10^6$, $B(0)=2\cdot 10^5$.
}
\end{figure}
\par
Figure 9 is connected again to the situation of slowly decreasing values of
the parameter $p$ with increasing time. In Figs. 9a - 9d this decrease is the
same. The differences among the figures 9a - 9d are because of the differences
in the values of the parameter $q$. The smallest value of $q$ is for Fig. 9a and
the largest value of $q$ is for Fig. 9d. The values of parameters in Fig. 9a
lead to a non-oscillation regime for the both substances in the cell of the
network. The increase of conversion rate $q$ leads to arising of a regime of
larger oscillations - Fig. 9b and with further increasing value of $q$ one
observes regime of smaller amplitude oscillation of the amount of the
substances that is followed by a regime of larger amplitude oscillations - Fig.
9c. The increasing of the value of parameter $q$ above a threshold value
leads to vanishing of the oscillation regime. Thus for the case of slowly
decreasing of the value of $p$ in the time the effect of the increasing $q$
is opposite to the effect of increasing $r$. 

\section{Application of obtained results  to the case of motion of migrants in a migration channel}
Let us now consider the models discussed above from the point of view of: (i) 
dynamics of motion of migrants in a migration channel and (ii) 
dynamics of migrant population $B$ and native population $A$ in 
a corresponding node (country) of the network of countries.  
Let us remember that from the point of view of 
 modeling of migration flows we shall consider the 
channel described in the previous sections as a chain of countries. Migrants
enter the channel from the entry country and may move through the
channel to its last node (the last country called final destination
country). In the general case  migrants may move in the both
directions: towards the final destination country or towards the
entry country of the channel. The ''leakage" in the channel is
connected to change of the status of some migrants, e.g., they
may obtain permission to stay in the corresponding country. 
This leakage may lead to situation in  which one may observe a 
presence of native population $A$ and population of migrants $B$ 
in a country. These populations have growth rates
($p$ and $r$).  In general $p$ and $r$   depend on the time.
In addition a process of integration may exist: the migrants of the population $B$ become
native citizens in the course of the time (case of positive values of the
parameter $q$). $q$ may have also non-positive values, i.e., the migrants
are not integrated and they may start to convert the native population 
that accepts the characteristics of the migrant population. Then  instead of 
integration one may observe absorption of the native population by the population of migrants.
\par 
Let us first discuss Figs. 2 and 3 from the point of view of migration dynamics. We note that for the
case of migration channel and for distributions where $f_i = \alpha_i + \beta_i i$, the parameters $\alpha_i$
characterize permeability of the the borders between the $i$-th and $i+1$-st country of the channel.
The parameters $\beta_i$ characterize the attractiveness of the $i$-th country of the channel and the
parameters $\gamma_i$ characterize the part of migrants that obtain permission to stay in the
$i$-th country of the channel. For the cases visualized in Fig. 2 the "leakage" parameter $\gamma_i$
has the same values for all four figures and for all countries of the channel.  
For the situation corresponding to  Fig. 2a the permeability of the borders decrease
with increasing $i$ and the attractiveness of the countries of the channel is the same. The form of
the distribution (\ref{dstr_main1}) is similar to the standard form of such distribution for the case of
infinite channel (see Appendix A). Specific effect connected to the finite channel arises in Fig. 2b.  The situation there is
characterized by increasing permeability of the borders between the countries of the channel with
increasing value of $i$ and by lack of attractiveness of the countries of the channel ($\beta_i=0$).
In this case the probability connected with the final destination country is larger than
the probabilities connected to countries of the channel with smaller value of $i$. This corresponds
to concentration of migrants in the final destination country of the channel. Let us call this effect
CP effects - effect of \textbf{C}oncentration because of \textbf{P}ermeability  of the borders. Fig. 2c shows
that CP effect can exist even if the attractiveness of the countries decreases with increasing value of $i$. Fig. 2d
shows that CP effect can lead to unusual form of the distribution of migrants where the probability increases with
increasing $i$ (this is opposite to the usual cases shown in Fig. 2a where probability decreases with increasing
value of $i$).
\par 
Fig. 3 demonstrates another kind of effect we shall call CA effect - effect of  \textbf{C}oncentration 
because of \textbf{A}ttractiveness. In all figures 3a, 3b, 3c, and 3d the value of the "leakage 
parameter" $\gamma_i$ is the same for all countries of the channel. The same is the situation with the 
value of the parameter $\alpha_i$. From Figs. 3a and 3b we observe the CA effect - if we set
the attractiveness parameter $\beta_i$ to $0$ then the large probability for the last country 
of the channel vanishes - Fig. 3b. This clearly shows the existence of CA effect. Thus we have two effects that influence the number of the migrants in the last country
of the channel (the final destination country) - CP effect and CA effect.  
Fig. 3d shows that CA effect can be quite large if the parameter $\beta$
has large enough positive vales.
\par 
Figures 4 - 9 describe the dynamics of native and migrant populations in a country 
in presence of inflow of migrants from a migration channel. We note that this situation 
may become more actual in the course of years as the human population of Earth still
increases and there are numerous military conflicts accompanied by poverty in many countries. 
In addition the climate changes may lead to large additional migration ("climate" migration).  
Above we have described 4 scenarios and let us now discuss these scenarios
from the point of view of dynamics of human migration. Fig. 4 shows the  effect of 
integration of the migrants in the corresponding
society. If the integration politics is consequent it can lead to increase of the 
native population in the course of the time even if the
rate of increase of the native population (accounted by the parameter  $p$) are much smaller 
than the rate of increase of the migrant population
(accounted by the parameter $r$).  Of course the capabilities of integration 
are limited. If the inflow of migrants is large and the
integration is not effective it is a matter of time for the situation to 
happen in which the number of non-integrated migrants will be larger and then
much larger that the number of people from the native population. 
\par 
Fig. 5 describes another possible scenario connected to a cyclic behavior of the number 
of individuals from native and migrant populations. This situation may arise when the 
rate of increase of the native population (births minus deaths) is negative and the rate of
conversion is positive (a number of migrants are integrated and become part of the native 
population). The period of the cycle depends on the rate $r$ of increase of the population of migrants. 
With increase of $r$ the period of domination of migrants population (the time in which the number of migrants is large than the number of individuals from the native population)  decreases . 
What we observe is a possibility of large intervals of time characterized by dominance of 
the migrant population. It may happen that the sign
of the parameter $q$ is reversed at some moment of such an interval. 
Then the native population of the country may become extinct.
Another scenario is shown in Fig. 6. Here despite  (i) the increasing inflow of migrants from the migration channel
and (ii) a positive values of the parameter $r$, stable dominance of the native population exists. 
The reason for this is the integration politics leading to positive value of the parameter $p$.
\par 
An interesting effect of negative value of the parameter $q$ (case when migrants integrate 
the native population) and negative value of parameter $r$ (the rate of increasing of migrants 
(births minus deaths) is negative) is shown in Fig. 7. When $r$ has small
negative values the interval of time of  domination of the migrant population can be quite large - 
Fig. 7a. When parameter $r$ decreases however the native population can become dominant despite the 
situation with the integration and a cyclic behavior occurs for the numbers of
populations of migrants and native individuals. 
\par 
The time-dependence of the parameters of the model of native and migrant populations leads to 
more complicated dynamics. Two possible scenarios are shown in Figs. 8 and 9.  We note the 
following: there is a specific diagonal in all cases of Figs. 8 and 9. This diagonal starts from 
the bottom left corner of the corresponding figure and ends at the top right corner of the figure.
If the phase trajectory is below this diagonal then there is a dominance of the native population 
(the number of individuals from the native population is larger than the number of individuals 
of the migrants population). If the phase trajectory is above this diagonal then 
the population of migrants is dominant. In Fig. 8a we observe a regime of small oscillations 
that happens in the area of dominance of migrants population.
The increase of the value of parameter $r$ leads to substitution of this regime by regime of
large amplitude oscillations - Figs. 8b, 8c. Finally a regime of dominance of population $A$
occurs - Fig. 8d. Fig. 9 shows several possible situations for the case when the rate 
$p(t_k)$ of increasing of the native population decreases slowly and in addition 
the rate of increasing of the migrant population increases from Fig. 9a to Fig. 9d. There is 
a constant rate of conversion of migrant population to native population.  The final results  
from this situation may vary: domination of the native population - Fig. 9a; 
large oscillations of the number of individuals with exchange of domination - Fig. 9b;
small oscillations with dominance of migrant population - Fig. 9c; 
final domination of the native population - fig. 9d.   
\section{Concluding remarks}
In this article we discuss a discrete-time model  of motion of substance in a finite-size channel 
of a network. The mathematical form of the general model is given by Eqs. (\ref {eq1}) - (\ref{eq3}). 
The particular case where the exchange of  substance between the channel and the environment happens only through 
first node of the channel is studied in more detail. For the stationary regime of motion of a 
substance through the channel we obtain a new class of statistical distributions  that contain as
particular case the truncated Waring distribution. Further we study the dynamics of two substances in a node of 
network that has access to the channel and because of this some amount of substance leaks from the channel to the 
studied node of the network. The second substance is specific for the node of the network and can interact with the 
substance that comes from the channel.  Four possible scenarios for the dynamics of the amounts of substance are 
described. The obtained results for the general case of motion of substance through
the channel are applied to the case of motion of migrants in a migration channel  that is positioned in a network of countries. Finally
in Appendix A we obtain the class of statistical distributions for the case of stationary motion of substance in a channel of infinite
length. These distributions contain as particular cases the famous long tail  Waring distribution, Yule-Simon distribution and Zipf distribution.
\begin{appendix}
\section{Statistical distribution of the substance for the case of channel
of network containing infinite number of nodes}
Let us consider the system of model equations for the case of infinite channel
that corresponds to the system of model equations (\ref{ey1})-(\ref{ey3}) for
the case of finite channel.  This system is
\begin{equation}\label{ew1}
x_0(t_{k+1}) = x_0(t_k) + \sigma x_0(t_k) - f_0 x_0(t_k) - 
\gamma_0 x_0(t_k)
\end{equation}
\begin{equation}\label{ew2}
x_i(t_{k+1}) = x_i(t_k) + f_{i-1}x_{i-1}(t_k) - f_{i} x_i(t_k) - \gamma_{i}x_i(t_k) \ \   
i=1,2,\dots
\end{equation}
Let us now consider a stationary state: $x_i(t_k) = x_i^*$. This
stationary state occurs when $x_i(t_{k+1})=x_i(t_k)$. From the system of
equations (\ref{ew1}), (\ref{ew2}) we obtain ($i=1,2\dots$, $x_0^*$ is a
free parameter)
\begin{eqnarray}\label{dstr2}
x_i^* = x_0^* \prod \limits_{j=1}^i \frac{f_{j-1}}{f_j+\gamma_j}
\end{eqnarray}
The total amount of the substance in the channel is
\begin{equation}\label{tasx}
x^* = x_0^* \Bigg[ 1 + \sum \limits_{k=1}^{\infty} \prod \limits_{j=1}^k 
\frac{f_{j-1}}{f_j+\gamma_j} \Bigg]
\end{equation}
We can consider the statistical distribution of the amount of substance along
the nodes of the channel $y^*_i=x_i^*/x^*$. For this distribution we obtain
\begin{eqnarray}\label{dstr_main}
y^*_0 &=& \frac{1}{\Bigg[ 1 + \sum \limits_{k=1}^{\infty} \prod \limits_{j=1}^k \frac{f_{j-1}}{f_j+\gamma_j}
\Bigg]} \nonumber \\
y_i^* &=& \frac{\prod \limits_{j=1}^i \frac{f_{j-1}}{f_j+\gamma_j}}{\Bigg[ 1 + 
\sum \limits_{k=1}^{\infty} \prod \limits_{j=1}^k \frac{f_{j-1}}{f_j+\gamma_j}
\Bigg]}; i=1,2,\dots 
\end{eqnarray}
Eq.(\ref{dstr_main}) describes a class of statistical distributions ($f_i$ and
$\gamma_i$ are still not specified). To the best of our knowledge this general
form of the class  of distributions was not discussed by other authors. 
\begin{figure}[!htb]
	\centering
	\includegraphics[scale=.7]{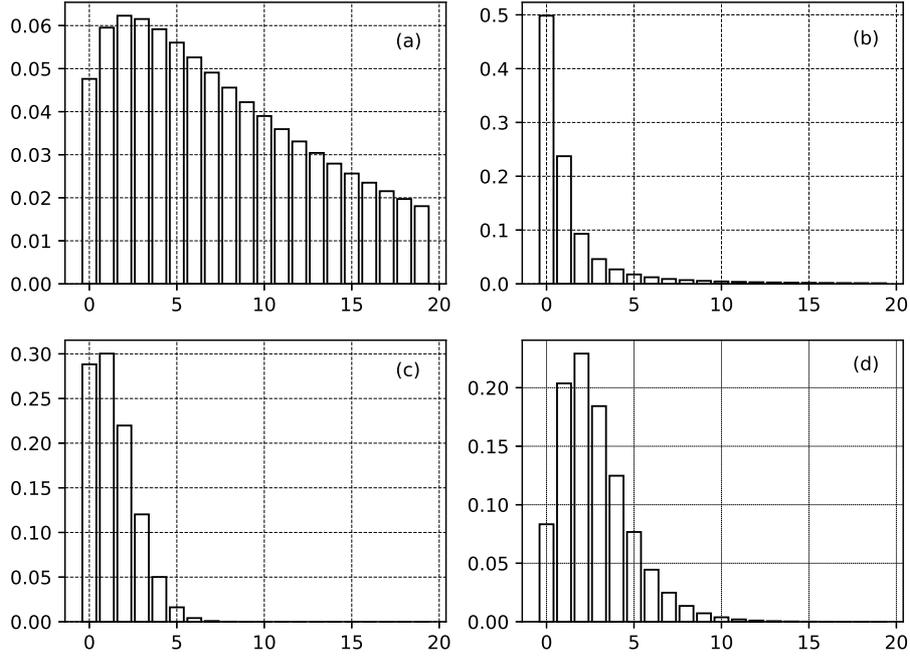}
	\caption{ The distribution (\ref{dstr_main}) . $f_i$ is selected to be of the kind: $f_i = \alpha_i + i \beta_i $ 
	where $\alpha_i$ and $\beta_i$ in  general can depend on $i$.  Figure (a): $f_i = 0.001/(i+1) + 0.001$, $\gamma_i = 0.0001$. The effect of decreasing $\alpha_i$ with increasing $i$ is shown. 
Figure (b): $f_i=0.001+0.001i^2$, $\gamma_i =0.001$. The effect of increasing $\beta_i$ with increasing $i$ is shown. Figure (c): $f_i=0.001/(i+1) + 0.0015$, 
$\gamma_i=0.001(i+1)^2$. The effect of decreasing $\alpha_i$ and increasing $\gamma_i$ is shown. Figure (d): $f_i=0.001/((i+1)^2) + 0.0001$, $\gamma_i=0.0001$. The effect of fast decreasing $\alpha_i$ with increasing $i$ is shown.}
\end{figure}
Fig. 10 shows that the shapes of the distribution depends on the form of $f_i$
and $\gamma_i$. $f_i$ is of the kind $f_i = \alpha_i + i \beta_i$. This can be
interpreted as $\alpha_i$ accounting for the permeability of the edge between $i$-th and $i+1$-th node of the channel and $\beta_i$ accounting for the ''attractiveness'' of the $i$-th node for the substance. Fig.10a shows that the decreasing
permeability of edges of the channel may lead to increase of the amount of 
substance around the entry node and then the amount of substance decreases 
smoothly in the depth of the channel. When the permeability decreases very
fast - Fig. 10d then the substance may concentrate in several nodes 
close to the entry node of the channel. Figure. 10b shows the influence of 
increasing '' attractiveness'' on the shape of the distribution: we observe
the form of a standard long tail distribution. Figure. 10c shown the effect of
decreasing permeability and increasing $\gamma_i$ which accounts for the
''leakage'' of substance in the corresponding node of the channel. 
\par
We note that the class of distributions (\ref{dstr_main}) has interesting
particular cases that have been discussed in connection with channels of
migration of substance or migration channels of human migration. For and
example let us consider the case where $\alpha_i$ and $\beta_i$ don't depend
on $i$, $f_i=\alpha_i+\beta_i i$,
$i=1,\dots,N$, $\alpha_i>0$, $\beta_i \ge 0$, $\sigma_0 >0$, $\gamma_i \ge 0$ 
(distributions of this kind have been discussed in \cite{sg1} - \cite{vb}) . 
For this case  the stationary distribution $x_i^*$ of the substance along the modes of the
channel is given by the relationship
\begin{eqnarray}\label{distr1}
x_i^* &=& \frac{\prod \limits_{j=1}^i [\alpha_{i-j} + (i-j) \beta_{i-j}] }{
\prod \limits_{j=1}^i (\alpha_j + j \beta_j + \gamma_j)} x_0^*, \
i=1,2,\dots \nonumber\\
\end{eqnarray}
and the statistical distribution connected to this stationary state of
functioning of the channel is
 \begin{eqnarray}\label{distr2}
y_0^* &=& \frac{1}{1+ \sum \limits_{i=1}^{\infty} \frac{\prod \limits_{j=1}^i [\alpha_{i-j} + (i-j) \beta_{i-j}] }{
\prod \limits_{j=1}^i (\alpha_j + j \beta_j + \gamma_j)} } \nonumber \\  
y_i^* &=& \frac{\frac{\prod \limits_{j=1}^i [\alpha_{i-j} + (i-j) \beta_{i-j}] }{
\prod \limits_{j=1}^i (\alpha_j + j \beta_j + \gamma_j)}}{1+ \sum \limits_{i=1}^{\infty}\frac{\prod \limits_{j=1}^i [\alpha_{i-j} + (i-j) \beta_{i-j}] }{
\prod \limits_{j=1}^i (\alpha_j + j \beta_j + \gamma_j)}}, \
i=1,2,\dots 
\end{eqnarray}
The distribution (\ref{distr2}) is a generalization, e.g., of the Waring 
distribution \cite{vk} as well as a generalization of one of
distributions discussed in \cite{vk1}.
\begin{figure}[!htb]
	\centering
	\includegraphics[scale=.7]{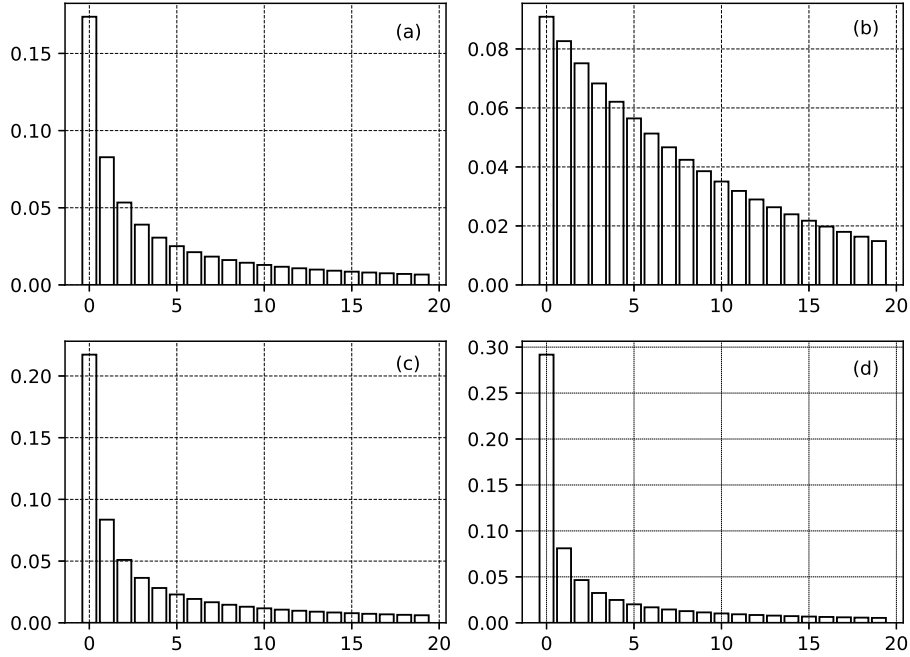}
	\caption{ The distribution (\ref{distr2}) for a channel of infinite length.
		All parameters $\gamma_i$ have the same value $\gamma_i=0.0001$ ($i=0,\dots,$) in all
		figures - 2a, 2b, 2c, 2d. All parameters $\alpha_i$ have the same value $\alpha_i=0.001$ ($i=0,\dots,$)  in all
		figures - 2a, 2b, 2c, 2d. Figures show the changes in the form of the distribution when the parameter $\beta_i$
		is changed.  Figure (a): $\beta_i = 0.001i$. Note that in the case of infinite channel the effect of concentration 
		of the substance in the last node of the channel is missing.
		Figure (b): $\beta_i=0$. The amount of substance in the nodes of the network decreases smoothly  from node to
		node of the channel.  Figure (c): $\beta_i=0.0015i$. $\beta_i$ are larger in comparison to Fig 2a. The effect
		of these larger values of $ \beta_i$ is concentration of substance in the entry node of the channel. This effect
		is seen also in Figure (d) where the values of $\beta_i$ are even larger: $\beta_i = 0.0025i$.  }
\end{figure}
Several forms connected to this distribution are shown in Fig.11. As we can see the changes in the values of the 
parameters $\beta_i$ influence the form of the distribution. From the point of view of application of theory to the 
case of channels of human migration parameters $\beta_i$ account for the attractiveness of the corresponding country 
of the channel. When all parameters $\beta_i=0$ (equally attractive countries) then the values of the distribution 
decrease slowly with increasing $i$. For the case of nonzero values of $\beta_i$ we observe increasing of the rate of 
migrants in the entry country of the channel. This effect is connected with the infinite size of the channel.
If the chanel has a finite size then a concentration of substance in the last node of the channel could be observed.

\end{appendix}
\end{document}